\documentclass{article}
\usepackage{amsfonts}

\usepackage{graphicx}
\usepackage{amsmath}

\newtheorem{theorem}{Theorem}
\newtheorem{acknowledgement}[theorem]{Acknowledgement}

\newtheorem{definition}[theorem]{Definition}

\newtheorem{proposition}[theorem]{Proposition}

\input{tcilatex}

\begin{document}

\title{Modular Theory and Eyvind Wichmann's Contributions to modern Particle
Physics Theory}
\author{Bert Schroer \\
Institut f\"{u}r Theoretische Physik der FU-Berlin\\
presently: CBPF, Rua Dr. Xavier Sigaud, 150\\
22290-180 Rio de Janeiro, RJ, Brazil\\
email: schroer@cbpfsu1.cat.cbpf.br}
\date{June 1, 1999\\
To be published in a special Festschrift volume, dedicated to Prof. E.
Wichmann on the occasion of his seventieth birthday }
\maketitle

\begin{abstract}
Some of the consequences of Eyvind Wichmann's contributions to modular
theory and the QFT phase-space structure are presented.

In order to show the power of those ideas in contemporary problems, I
selected the issue of algebraic holography as well as a new nonperturbative
constructive approach (based on the modular structur of wedge-localized
algebras and modular inclusions) and show that these ideas are recent
consequences of the pathbreaking work which Wichmann together with his
collaborator Bisognano initiated in the mid 70$^{ies}.$
\end{abstract}

\section{Wichmann's Influence on Ideas in Local Quantum Physics}

Looking at the various contributions to post perturbative quantum field
theory, one could try to group ideas, which shaped the form of our present
understanding, into three classes. There are first those ideas on
nonperturbative frameworks whose's usefulness was immediately obvious. The
prime example would be the time dependent scattering theory of Lehmann,
Symanzik and Zimmermann\cite{WeiB}\footnote{%
Since I did not intend this essay to become a review article, I will use
texbook references whereever possible (from which the reader may find an
exposition of the content of the referred articles as well the reference to
the original paper(s)).} which together with the resulting stationary
scattering formulas and combined with the dispersion theoretical framework,
marked the beginning of at least the (kinematical) setting of
nonperturbative model-independent thinking. This theory (or rather this
framework) used the formalism of (interpolating) local fields and their
correlations. Fortunately there already existed at the time of start of LSZ
a mathematically as well as conceptually very concise setting for doing
quantum field theory, namely the characterization (of what used to be from
the outset an operator theory) in terms of Wightman functions which gave an
excellent understanding of the general singular function nature of the
correlation functions, as well as of their analytic continuation aspects 
\cite{St-Wigh}. The conceptually important aspect of that framework was of
course the reconstructability of operators, states and Hilbert spaces from
those functions; physicists traditionally, since the times of Feynman's
great contributions, prefer to deal with functions and could not have been
completely sure that their description is complete. The addition of the LSZ
formalism also bridged the gap between Wightman's pure field theory setting
and the 1939 theory of one-particle spaces as irreducible representation
spaces of the Poincar\'{e}-group. This connection of field theory with
particle theory was later amplified in work by Weinberg \cite{WeiB} which
linked Wigner's particle theory with Feynman's perturbation theory (in fact
the exposition of the Wigner theory, the historical remarks as well as the
presentation of the QED calculations constitute the high points of
Weinberg's textbook). From there on it was clear that QFT, in contrast to
classical field theory (viz. the particle models of the electron of Poincar%
\'{e} and Lorentz), did already contain in principle all particle (including
multi-particle scattering) aspects, and even more, that there were methods
and beautiful formulas which explicitly allowed to extract those particle
properties of fields.

In a way the Wigner theory of relativistic free particles was the first
successful attempt to present relativistic particles without referring to
quantization (in this case of relativistic classical mechanics) and in this
way it was doing justice to the more fundamental nature of quantum over
classical theory. Because of this, Wigner's theory became exemplifying for
all attempts to formulate and execute problems of local quantum physics
(LQP) without relying on the quantization of classical expressions as
Lagrangians. One of its immediate successes was the classification of the
plethora of physically equivalent field equations, which physicist found in
the aftermath of Dirac's discovery for the description of relativistic
electrons/positrons.

This trend of finding more intrinsic descriptions for local quantum physics
was one of the motives behind the algebraic approach initiated by Haa, which
reached its first stage of mathematical maturity (``the framework of
Algebraic QFT'') in a often cited paper of Haag and Kastler. The guiding
principle there was a vast generalization of the idea of Wick, Wightman and
Wigner on superselection sectors from their univalence rule in the direction
of superselected generalized charges. The emerging paradigm of algebraic QFT
was to view all of QFT, i.e. including the issue of spin and statistics and
scattering theory, as the synthesis of the superselected representation
theory of an (model dependent) underlying observable algebra with the
localization structure imposed by Einstein causality \cite{Ha}\cite{Bu}\cite
{SchAOP}

This program required a formulation which was independent on ''field
coordinatizations'' from the very beginning. Wightman's approach \cite
{St-Wigh} on the other hand was more conservative in that it used field
operators which in contradistinction to the neutral observables were allowed
to carry charge and to appear in multiplets acted upon by internal symmetry
groups with the effect that for most of the time it stayed closer to the
very successful perturbative Lagrangian QFT. The conceptual relation between
these two approaches was clarified by Doplicher, Haag and Roberts as well as
Borchers \cite{Ha}; their detailed mathematical connection offers some
challenges up to date. But even in the first comprehensive account of the
Wightman theory \cite{St-Wigh}, these authors already went some distance to
counteract the tendency of overemphasizing the role of fields by pointing
out that what is really relevant are equivalence classes (Borchers classes)
of relatively local fields.

While touching the issue of scattering theory, one should not leave
unmentioned another well-embraced idea of those early times: the S-matrix
bootstrap approach. This was mainly Chew's idea (with some prior attempts in
this direction by Heisenberg). In its extreme form of cleansing quantum
fields and the locality principle altogether from the arena of quantum
physics, it eventually failed. But, even apart from the useful pictures
about effective interactions (as they were later obtained by qualitatively
reading back certain on-shell properties into QFT by Weinberg), it left some
interesting structural elements behind (see later); which is really a lot
one can say as far as failed theories are concerned. In fact one may call it
the most successful among all failed theories of this century, a statement
whose content will only become comprehensible in the light of its relation
to modular theory sketched at the end of this essay.

The reason I mention these trends and achievements of the 50-70$^{ies}$ is
that, unlike most of his contemporaries in general QFT, Eyvind Wichmann, in
whose honor I wrote the present essay, did not enter general QFT directly
but rather started his carrier with very detailed QED radiative correction
calculations and came, after passing actively through other particle physics
problems, permanently to Berkeley as a result of his active interest in
S-matrix theory and the dispersion relation approach.

But now it is time to get to the idea which is inexorably linked with the
name of the Jubilar\footnote{%
In trying to translate the German ``Jubilar'' into short and precise
English, my dictionary said: ``the person whose anniversary is being
celebrated''. This is precisely the meaning, but not the type of short
expression I was looking for.}, namely the observation that QFT localization
is related in a very deep way with the fundamental mathematical modular
theory for von Neumann algebras.

The first achievement in this direction culminates in a 1975 some seminal
paper by Wichmann and (his at that time Ph.D. student) Bisognano \cite{Ha}.
It relates Tomita's 1967 modular theory, dealing with basic structural
properties in von Neumann algebras, with fundamental structures of
nonperturbative QFT\footnote{%
Nowadays mostly referred to as the Tomita-Takesaki modular theory because it
was M. Takesaki, who by his penetrating analysis and improvements of
Tomita's result, contributed to its widespread acceptance within the
community of operator algebraist (and finally led A. Connes among other
things to his classification of type III von Neumann factors).}. Already at
the time of Tomita's presentation of his theory at the 1967 Baton Rouge
conference in the US, there was a physics discovery which preempted certain
aspects of Tomita's theory. This was the analysis by Haag, Hugenholtz and
Winnink of thermal quantum physics directly done in the so called
thermodynamic limit of an infinite extended QFT\footnote{%
When the standard box-quantization produces a conceptual clash with other
ideas, as it is the case for e.g. (time-dependent) scattering theory or
phase transition properties in statistical mechanics, it is clearly
preferable to understand the infinitely extended translationally invariant
systems directly and to view finite systems as open subsystems.}. Their
crucial observation consisted in noticing that a calculational trick, which
appeared in previous works of Martin, Kubo and Schwinger (refereed to as the
KMS condition) takes on fundamental conceptual role in the physics of
translational invariant local statistical quantum systems \cite{Ha}. Whereas
Winnink immediately after the Baton Rouge conference, where both results
were presented independently, began to elaborate the deep connections in his
thesis with Hugenholtz, Haag and his collaborators succeeded to derive the
KMS condition from the stability properties of statistical mechanics
equilibrium states \cite{Ha}. This line of thinking culminated in the work
of Pusz and Woronowicz \cite{Ha} by which the abstract Tomita modular
theory, if enriched by the physical idea of locality, got directly linked
with the second fundamental law in thermodynamics, i.e. the impossibility to
construct perpetuum mobiles and all that. As a corollary, also the TCP
symmetry of local QFT develops another ``modular'' relation (in addition to
the ``detailed balance'' relation) to equilibrium statistical mechanics and
in particular to the second fundamental law.

So at the time when Wichmann with his collaborator Bisognano discovered the
connection between wedge localization of quantum fields and the modular
objects of Tomita for this situation\footnote{%
In mathematical terms the problem was to compute the modular objects
(modular group and modular involution) for the algebra generated by quantum
fields smeared with functions which have their supports in a wedge within
the vacuum representation. The modular group turned out to be the wedge
associated Lorentz boost and the modular involution was the TCP-like
antiunitary reflection along the ridge of the wedge which maps the wedge
into its Einstein-causal complement. Certain very special free field modular
localization aspects where already noticed before\cite{Sch}.}, some of the
thermal (heat-bath) aspects were already well understood. It is very natural
that this situation called for an analysis of the Hawking-Unruh effect in
those modular terms. But the first paper in this direction was a
contribution by G. Sewell \cite{Ha}.

Another line of research of Wichmann was his collaboration with D. Buchholz
who, being familiar with prior work by Haag and Swieca, brought Wichmann's
attention to this problem; the fruits of the joint discussions finally led
to further significant contributions to the clarification of the degrees of
freedom or phase space structure problems in QFT \cite{Ha}. As many
analogies of QM with LQP did not persist under closer conceptual scrutiny,
the clarification of the nature of LQP phase space was an important issue.
Already noted in the seminal work of Haag and Swieca \cite{Ha}, the counting
with a ``relativistic box'', i.e. for a Minkowski space double cone region
together with a sharp energy cutoff, could not give a finite number of
degrees of freedom per phase space cell as in the case of the box in QM.
Their rather rough methods and estimates were improved in the
Buchholz-Wichmann work and it became clear that the correct counting could
not be better than ''nuclear''\cite{Ha}, not even in the interaction free
case. It also gave a deeper insight into a prior conjecture that quantum
field theory should exhibit the ``split property'' which is the statement
that by allowing a ``collar'' region between the inside and the outside
world of a double cone (which is the QFT counterpart of the quantum
mechanical space box), the total algebra allows a tensor factorization of
the inside and outside algebras which is not possible without that collar.
Later it turned out that this structure also had a deep relations with
modular theory. A third line of research of the Jubilar, which probably
brought him to Berkeley, was on S-matrix theory. Here I am in the
comfortable position of being able to refer to Weinberg's book where justice
is done to Wichmann's S-matrix articles \cite{WeiB}.

Wichmann's contributions definitely belong to those ideas which, unlike e.g.
LSZ, did not have a visible connection to the immediate problems of particle
physics nor could one draw upon them as a framework in QFT\ as with the
aforementioned contributions of Wightman as well as those of Haag, Kastler,
Jost and others. Like most of Borchers contributions, they were less of a
systematic encyclopedic but rather more of an enigmatic nature. Their power
only unfolded slowly with time, and even up to date we are still witnessing
an accelerating unfolding process of the physical consequences of modular
structures; a fact which probably even Wichmann did not expect when he
investigated this subject in the mid 70$^{ies}.$

The enigmatic power of Wichmann's modular ideas has been brought to light in
many articles ranging from thermal QFT to superselection structure i.e. the
reconstruction of charge-carrying fields including their statistics,
symmetry properties and TCP structure which especially in low dimensions
(left out in Wightman's framework) gives rise to very new and nontrivial
problems related to modular theory \cite{FRS}\cite{Lo}\cite{Bo}\footnote{%
Here the interested reader may also consult my recently 62 page review
article (also containing additional more recent references) which I
dedicated to Eyvind Wichmann and accepted for publication in JPA, but which
I could not reproduce in the special Festschrift volume because of copyright
laws.}.

Rather than continuing the flow of history, I would like to try in the next
section to exemplify the power of modular ideas in the context of two very
recent (and still unfolding) ideas: holography and a new nonperturbative
method (which still lacks a known catchy name) based on the structure of
those Bisognano-Wichmann wedge algebras.

Since holography, in the sense of encoding the content of a QFT into a lower
dimensional one has, especially in the context of QFT in AdS (Anti-deSitter
spacetime) attracted a lot of attention, this may be a good vehicle for
demonstrating the power of modular ideas to a broader QFT knowledgable
public. This is particularly the case in view of the perplexing aspects
which holography and in particular quantum matter in AdS presents to the
standard quantization (Lagrangian or functional interaction) approach. Most
of these paradoxical aspects are naturally solved by algebraic QFT with the
modular enrichment originating from the Bisognano and Wichmann work being
the essential link between local quantum physics and geometry of AdS \cite
{Re} \cite{BFS}. These are methods and concepts which are not yet known
outside a small circle of specialists, although the latter paper makes a
great efforts to use rigorous physical arguments and avoids the explicit
evocation of modular theory (but rather derives it, because all the
arguments are of a general QFT nature). And even if, after all, the
conceptual barriers for many particle physicists will remain high (please
dear reader, remember that you did not learn the sophisticated differential
geometry and topology many of you are using these days in less than a
year!), their enthusiasm for the AdS-conformal QFT connection may be of help
here.

The constructive approach via the modular properties of wedge algebras which
is taken as our second illustration (which I initiated and pursued over the
last years, partially with my collaborator H-W Wiesbrock) \cite{Sch}\cite
{S-W1}\cite{SchGoe}, is best thought of as the inverse of the
Bisognano-Wichmann theorem, namely using the modular theory for the actual
construction of \ quantum field theories. It shares with holography issue
the fact that a real understanding inside any framework based on
``field-coordinatizations'', in particular in any quantization approach, is
impossible\footnote{%
At least if one does not solve the Lagrangian theory nonperturbatively and
explicitly, and then reprocesses the resulting dynamical variables; an
altogether impossible task.} (or only possible with a mind which has the
power of artistic imagination). Here one is really hitting the limits of
capabilities of what the quantization access to QFT can do, and although
analogies are always dangerous, I feel tempted to make a historical
comparison with Bohr-Sommerfeld rules (please don't look down on them, they
where quite successful for many problems, but they demanded some artistry)
versus full quantum mechanics.

Indeed everything you learned in text books (and which proved so useful in
the pursuit of perturbation theory) about standard quantum field theory, as
interaction picture, time ordered correlations, canonical quantization or
quantization through euclidean functional integrals, all that is of no avail
here. Of course non of these well-known standard tools is typical for LQP%
\footnote{%
The reader may have noticed that whenever we want to emphasize the concepts
of QFT\ but not necessarily their present textbook implementations, we
prefer the termonology LQP (local quantum physics). This diminishes the
unwarranted (almost subconscious) tendency of the reader to equate QFT with
euclidean funcional representations and all that.}, all of them can (and
have been) applied to QM as well, either in its Schr\"{o}dinger or its
so-called second quantized form. \textit{But the modular structure in QFT on
the other hand is totally characteristic} and not shared by any other kind
of quantum theory. 

Anybody who is familiar with the conceptual framework of LQP knows that this
is deeply related to Einstein causality and the polarization structure of
the vacuum resulting from that causality in the presence of interactions,
the denseness of localized states generated by acting with operators
associated with nonvanishing region\footnote{%
This property has been colloquially termed the ``particle behind the moon''
paradoxon, better known as the Reeh-Schlieder property or in mathematical
language of von Neumann algebras as the cyclic and separating property of
the vacuum with respect to local algebras.} whose causal complement is
nontrivial, or the total different nature of local algebras as compared to
algebras of QM with consequences for quantum measurement theory \cite{SchGoe}%
. Especially these last structural differences may be somewhat surprising
(in the sense of low credibility) to somebody who has always thought about
QFT just as a relativistic continuation of QM. Hyperfinite type III$_{1}$
von Neumann factors, as the wedge localized algebras of QFT, have indeed
very different properties from the type I algebras of QM. They do not permit
pure states at all\footnote{%
We follow here the by now standard terminology to omit the prefix ``normal''
for states, and add the prefix singular for the rare case of nonnormal
states on von Neumann algebras.}, and they also, together with their
commutants, do not admit the notion of tensor product factorization of the
total algebra B(H), disentanglement and all the other old notions from von
Neumann's exposition of QM, which recently have been propped up for
''quantum computation'' \cite{SchGos}. In order to recover the usual quantum
mechanical structure of the inside/outside factorization of a Schr\"{o}%
dinger box, one has to work quite hard and use the Buchholz-Wichmann
nuclearity property for the control of local degrees of freedom \cite{Ha}.
The resulting ``relativistic box'' consists of a the inner region of a
smaller double cone and the outer (infinite) spacetime of a larger double
cone, but note that the inside and the outside needs to be separated by a
``collar region'' in order to attenuate the uncontrolled vacuum fluctuation
caused by sharp boundaries (the latter trouble already having been known to
Heisenberg in his study of vacuum polarizations). As mentioned before, this
is known under the name of split situation or split inclusions.

There are many recent results from modular theory which all point into the
same direction and contain the same general message, namely one is dealing
here with structures which, if at all, only with a superhuman hindsight and
extraordinary stretch of imagination are visible from a quantization
framework. Another convincing illustration not presented here, are the
recently found ``hidden symmetries'' \cite{S-W2}\cite{BDFS}, where the word
hidden is used in the sense of hidden to the Lagrangian-Noether framework
(and of course not in the sense of the modular framework within which they
were discovered). We also refrain from presenting some very surprising
results about the possibility of creating a local net in LQP together with
the full Poincar\'{e} symmetry from just a few (for chiral conformal theory
2, for d=1+2 theory 3 and for d=1+3 theory 6 ) algebras in a certain modular
position to each other \cite{Wies}\cite{S-W}. Although algebraic QFT, unlike
string theory, is not designed to be quantum gravity, these findings about
totally unexpected relations between raw (highly noncommutative) algebraic
data and spacetime geometry, although not being directly related to quantum
gravity, should be taken serious in any attempt towards quantum gravity.

Another very important consequence of modular theory is the already
mentioned thermal aspect which it attributes to localization. In the case of
``natural localizations'' related to classical bifurcated Killing horizons
as they occur in black hole physics, this thermal aspect can, and as
everybody knows, has been discovered before modular theory. But for the
general localization in QFT, which cannot be describe in such classical
metric properties, one really needs it. The modular localized subspaces are
dense in the Hilbert space of the full theory, but there is a natural
``thermal'' scalar product (the graph metric defined by the unbounded Tomita
operator S of that region) in terms of which it is closed. This thermal
inner product changes with the localization region of the local algebras,
and it turns out to be related to the domain problems of Wightman's theory%
\footnote{%
Actually the Wightman domain is related to the intersections of all
(thermal) modular domains. This is quite interesting, since many particle
physicist in my generation were told not to worry to much about these domain
problems and accept them as a technical mathematical assumption void of any
direct physical interpretation.} and possibly also to the construction of
pointlike covariant fields from the net of local algebras. There is a
speculative remark of Fredenhagen (privat communication) which fits in very
nicely with these physical aspects of field domains and ranges of actions of
algebras on the vacuum. It is the idea that a pointlike field, or rather the
one-field subspace obtained by its application to the vacuum, can be
characterized as the carrier of an irreducible representation of some
(infinite dimensional) ``universal modular group''. The latter is generated
by all the one parametric modular groups for all spacetime regions\footnote{%
The dynamics would then be carried by the representations of the local
modular groups instead of the global time-frame dependent Hamiltonian.}.
This, if true (it is true for those QFT which have been by the wedge
localization method sketched in the next section), would make a rather
pivotal addition (if not revolution) in QFT as it has been hitherto
understood since it attributes to pointlike fields the an analogous
intrinsic physical role as the Wigner positive energy representation theory
of the Poincar\'{e} group. In this way the ``fields'' would recover some of
their lost ground (at least in the form of the mentioned field spaces), when
from the viewpoint of AQFT they became relegated to mere ``coordinates'' of
algebras. And much more: since the modular groups and their unitary
implementers are expected to contain the crucial information on interaction,
they would gain in addition to the geometric properties they already had in
the quantization approach, the status of an intrinsic modular-based concept
of interaction. To put it into the context of the more concrete constructive
nonperturbatve modular setting of the next section, the incoming particle
content of the interacting field (in terms of its formfactor spaces which
appear in their decomposition) would be governed by a new and subtle
(hidden) kind if infinite-dimensional group theory as a kind of analogue of
the (overt) diffeomorphism group in chiral conformal field theory\footnote{%
There are strong arguments that those chiral diffeomorphism groups are of
modular origin and belong to multi-interval algebras together with specially
chosen (non-vacuum) states \cite{S-W}.
\par
.}. The characterization of special operators in such an algebra would then
require the study of the relation to modular subgroups belonging to finer
localizations inside the chosen one.

\section{ Holography and the Constructive Approach to Wedge Algebras}

Holography is the conjectured correspondence between higher and (conformal)
lower dimensional QFT (or a family of lower dimensional ones). The
attractive aspect of such a correspondence is that a lot more is known about
low-dimensional QFT, in particular conformal QFT, which could be of use for
the construction of higher dimensional QFT's. Historically the idea can be
traced back to the thermal and geometric behavior of black hole (classical)
entropies (Bekenstein, Hawking). Since the temperature aspects were
understood in the setting of (free) QFT in CST, it was only natural to look
for an explanation of the surface proportionality of entropy in terms of
quantum degrees of freedom at or near the horizon. In contrast to the
understanding of the Hawking-Unruh temperature as originating from the
causal localization behind a (Killing) horizon, the entropy problem was less
susceptible to explicit calculation involving (free) quantum matter in black
hole background. But it is clear that if one could understand the surface
nature of the degree's of freedom, then the entropy should follow suit. In
the Lagrangian formulation of QFT the elusive ``light-cone physics''
preempted some aspects of this idea, and is not surprising that 't Hooft 
\cite{Ho}, who on various occasions used light-cone quantization, in more
recent times suggested to interpret Bekenstein classical observation on
black hole entropy in terms of quantum ``holography''.

A problem like the present one, where field coordinates are not transformed
into each other, but rather degrees of freedom become transmuted in a way
which is hard to describe in terms of pointlike field concepts, is bound to
cause trouble within the usual quantization formalism. To be sure, problems
with the use of one set of field coordinates versus another one already
appeared before in QFT, although in the early 60$^{ies}$ they were sometimes
the source of some prejudices about Lagrangian fields being in some sense
``better'' than any other composites (carrying the same charges). This was
part of a bigger confusion about particles versus fields; the elementary
versus bound state hierarchy of QM tacitly entered QFT where it should have
been replaced by the hierarchy of superselected generalized charges and
their fusion (including those nonabelian Casimir charges which underlies
nonabelian internal symmetries). For example in connection with the PCAC,
physicists in Lagrangian field theory had to take notice of the fact that
one is not slavishly bound to those field coordinates in terms of which one
has written a particular Lagrangian. From the time in Illinois, as a
collaborator of Rudolf Haag, I remember a conversation between Murray
Gell-Mann and Rudolf Haag which ended with some astonishment on the side of
Gell-Mann. Nowadays the understanding of the extreme insensitivity of
onshell objects like the S-matrix against changes of field coordinatizations
has become a commonplace even in Lagrangian QFT, especially after Weinberg
tought physicists how to formally handle this problem in perturbation
theory. However the morphisms and isomorphisms needed in order to understand
holography are of a different caliber.

Mature physicists are of course aware that progress in physics is to a large
part the liberation from prejudices (including ones owns). Algebraic QFT
theory had the big advantage that there was no place for prejudices about
fields, because there were no fields in its formulation. As a result, the
problem with this ``field coordinate free formulation'' was shifted
somewhere else. Namely it was not entirely clear that, although one did not
want to invent a new theory but just implement the same physical principles
which underlie the quantization approach in a different conceptually and
mathematically more controllable way, that one had not in fact actually lost
the connection to the same particle physics. But it became soon clear that
the gains of understanding by working with nets of operator algebras instead
of fields (as e.g. the manifest independence of the S-matrix from the choice
of particular interpolating operators taken from the local algebras) were
not offset by an undesired vagueness or unintended invention of new physical
content. This was established beyond reasonable doubt, and in the case of
chiral conformal QFT there is even a rigorous proof for the equivalence of
the Wightman description with the algebraic framework \cite{F-J}. As
mentioned before algebraic QFT based on modular methods wants to stay
laboratory physics, and not like string theory aim at quantum gravity.

Already in the early stages of the theory there were concepts, questions,
and techniques which transcended the Lagrangian framework and even that of
Wightman. One could e.g. ask about the possibilities of particle statistics
compatible with the Einstein causality of observables. This goes certainly
beyond the Wightman theory which is a theory which includes the
charge-carrying fields which ab inicio are assumed to have $\pm $
commutations relations for spacelike separation. The Spin-Statistics theorem
just selects the correct one of these two possibilities. In algebraic
quantum field theory one succeeds to compute the field statistics without
such restrictive assumptions on nonobservable quantities (on which in d%
\TEXTsymbol{<}1+3 one has anyhow no a priori control). In the intermediate
steps of the conceptually and mathematically rich DHR and DR constructions 
\cite{Ha}, there appear parastatistics fields which belong to nonabelian
Young tableaux of the permutation group. They do not permit quasiclassical
limits and Lagrangians, but are reasonable objects in algebraic QFT (in the
sense that the charge carrying parastatistics fields have enough locality in
order to admit a reasonable physical interpretation, albeit one which is
much more noncommutative. Only after enlarging the Hilbert space by the
introduction of multiplicities (i.e. indices on which symmetry-groups can
act), does one make contact with quasiclassics. On the other hand, writing
down a Lagrangian has already preempted the answer before having been able
to ask the question. An ardent philosophical empiricist may point to the
fact that there was never any practical need for asking such a question
since the usual formulation with built in multiplicities and Bose/Fermi
statistics works nicely. But he would have a hard time in say d=1+2 theories
with braid group statistics, where it can be mathematically demonstrated
that those plektonic objects will never fit into a Lagrangian quantization
approach with field multiplicities.

Of course such an empirical fundamentalist may then retort that d=1+2 models
is not particle physics. In that case one could, assuming that he does not
also declare the present AdS discussion in connection with holography for
irrelevant in particle physics, point out to him that although there is no
satisfactory solution of the paradoxical situation of holography in any
quantization approach\footnote{%
Quantization difficulties have been mentioned in Witten's papers \cite{Wi}.
Actually AdS only exposes the \textit{general limitation of quantization}
which always exists in any interacting QFT, once one leaves the realm of
perturbation and quasiclassical approximations. The entrance into QFT from
the noncommutative side of modular theory i.e. without the classical
parallelism (called quantization) may be more difficult and unusual, but
does certainly not suffer from those limitations.}, the solution which was
given in algebraic QFT by Rehren \cite{Re} is conceptually clear and
mathematically rigorous. The main point in Rehren's presentation is that the
adapted Bisognano-Wichmann theorem allows to understand an isomorphism
between AdS$_{n+1}$ and conformal Minkowski spacetime $M_{n}$ which is not a
pointwise geometric mapping (diffeomorphism) but rather a set mapping
between modular localization regions of algebras. In fact the notion of
``weak locality''\cite{St-Wigh} used in his paper is completely equivalent
to the spatial part (the thermal subspaces of the total Hilbert space which
are closed in the modular graph norm) in my constructive approach built on
modular wedge localization \cite{SchAOP}. If the reader finds the small
amount of modular concepts inaccessible because he lacks mathematical
understanding of LQP concepts, he may have another chance by looking at the
closely related paper of Buchholz, Florig and Summers \cite{BFS}. These
authors explain the LQP in an AdS space-time together with a rigorous
physical account of what is necessary to know about modular theory without
assuming (in principle) a prior knowledge. We will not try to reproduce
these results here, since the clarity of the papers makes this a sacrilege.

As far as I could see, the only open problem in the BFS work is the question
whether there can be any genuine interaction at all in such a AdS world with
that causality paradox mentioned in their paper. This question is
reminiscent of a problem which I encountered in my collaboration with Swieca
at the beginning of the 70$^{ies}.$ At that time there existed the challenge
to understand the (global) ``causality paradox of conformal QFT'' \cite{HSS}%
, i.e. the apparent contradiction between being able to conformally
transform oneself globally from space-like separations via the light-like
infinity into the time-like region and the fact that certain interacting
``would-be'' conformal models, as e.g. the massless Thirring model, did not
comply with the Huygens principle calling for vanishing time-like
(anti)commutators which was required in order to avoid contradictions with
that global transformation property. In fact the only known d=1+1 models
which did not generate this paradox were free fields with Fouriertransforms
on the light cone, as conformal currents or energy-momentum tensors. The
resolution \cite{S-S}\cite{Pal} of this paradox turned out to consist in
realizing the important role of the conformal covering space in that those
paradoxical looking fields as the Thirring field were not (as everybody
believed up to that time) globally irreducible, but rather had a rich
decomposition with respect to the center of the conformal covering group.
The irreducible components in this decomposition (there simply called
``nonlocal components'') became known 10 years later as the conformal blocks
in the famous Belavin-Polyakov-Zamolodchikov paper as everybody knows. One
reason why we only looked at the Thirring-like exponential boson fields was
(besides the fact that they already were available), that these new
irreducible component objects were outside the range of euclideanization and
even outside the Wightman framework\footnote{%
Wightman fields do not come with source and range projectors as those
nonlocal components, for an explicit illustration see \cite{R-S}.} since
their algebra admits local annihilators. My impression after having read 
\cite{BFS}is that the AdS situation has analogous causality problems. In
fact, using the Rehren isomorphism for AdS(1,1), one would expect to be able
to lift the solution of the old conformal paradox directly into the new
AdS(1,1) realm.

I now would like to explain some of the modular ideas which I used recently
in a constructive program for interacting LQP models which is based on the
use of algebras and is completely free of field coordinates (although I will
think of the reader, and use a field notation whereever possible). Of course
one must first test these ideas in the interaction free case. This I did by
showing that the Wigner representation theory can be directly used for the
construction of the local nets without e.g. using Weinberg's formalism of
first constructing free fields which then would generate these local
algebras. In this way one obtains an intrinsic description of noninteracting
theories which restores (or rather maintains) the uniqueness\footnote{%
The uniqueness on the field level was lost because there are infinitely many 
$u$- and $v$- interwiners from the unique Wigner representation to the
plethora of covariant L-representations. Weinberg prefers the one (those)
for which there is a free Lagrangians, since he wants to use fields for a
euclidean functional integral representation. Real time causal perturbation
theory on the other hand can be done in any field coordinatization. The best
way would be do use none at all.} of the (m,s) Wigner representations and
avoids the plethora of covariant associated field coordinatizations \cite
{Sch}\cite{S-W1}\cite{SchGoe}.

This first step may be viewed as analogous to the intrinsic coordinate-free
description of geometry. It uses a kind of inverse of the Bisognano-Wichmann
theorem, i.e. the known modular theory for the wedge, in order to obtain the
operator algebra localized in the wedge. It may be viewed as a refinement of
Weinberg's exposition of the Wigner theory mentioned in the first section,
by implementing the idea of modular localized subspaces directly in the
Wigner momentum space description without the use of covariant x-space wave
functions or the noncovariant Newton-Wigner localization. This baby-version
of modular theory can be understood without knowing anything about the
Tomita modular theory and as such furnishes an excellent pedagogical example
of the power of modular localization and the Bisognano-Wichmann theorem.

The next step namely to construct interacting nets in this intrinsic manner
is more difficult. Of course one could follow Weinberg for the construction
of free fields from Wigner particles, select some free fields corresponding
to (m,s), couple them to a scalar Wick-ordered interaction density W(x)
(which one may call $\mathcal{L}_{int},$ but the existence of an $\mathcal{L}%
_{0}$ is not necessary, see previous footnote) which is then plugged into
the causal perturbative machine whose heartpiece is the perturbative
transition operator $S(g)$. From there one obtains the retarded
representations of interacting fields in terms of free field in Fock space
which in turn (or by direct use of the $S(g)$) generate the localized
algebras after suitable test function smearing. But this way of constructing
local algebras would amount to just an exercise in semantics and go against
the spirit of LQP.

Let me explain the gist of the correct idea with the help of a
two-dimensional representation and using standard field theoretical language
wherever it is possible.

Let $A(x)$ be a d=1+1 massive free scalar fields with the following notation$%
.$%
\begin{eqnarray}
A(x) &=&\frac{1}{\sqrt{2\pi }}\int (e^{-ipx}a(p)+h.a.)\frac{dp}{2\omega } \\
&=&\frac{1}{\sqrt{2\pi }}\int (e^{-im\rho sh(\chi -\theta )}a(\theta
)+h.a.)d\theta ,\,\,x^{2}<0  \notag \\
&=&\frac{1}{\sqrt{2\pi }}\int_{\mathsf{C}}e^{-im\rho sh(\chi -\theta
)}a(\theta )d\theta ,\,\,\,\mathsf{C}=\mathbb{R\cup }\left\{ -i\pi +\mathbb{R%
}\right\} \mathsf{\,\,\,}  \notag
\end{eqnarray}
where in the second line we have introduced the x- and momentum- space
rapidities and specialized to the case of spacelike $x$, and in the third
line we used the analytic properties of the exponential factors in order to
arrive at a compact and (as it will turn out) useful contour representation.
Note that the analytic continuation refers to the c-number function, whereas
the formula $a(\theta -i\pi )\equiv a^{\ast }(\theta )$ is a definition and
has nothing to do with analytic continuations of operators\footnote{%
Operators in QFT never possess analytic properties in x- or p-space. The
notation and terminology in conformal field theory is a bit confusing,
because although it is used for operators it really should refer to vector
states and expectation values in certain representations of the abstract
operators. The use of modular methods require more conceptual conciseness
than standard methods.}.

With this notational matter out of the way, we now write down our Ansatz for
nonlocal but (as it turns out) still wedge localized fields using the same
notation 
\begin{equation}
F(x)=\frac{1}{\sqrt{2\pi }}\int_{\mathsf{C}}e^{-im\rho sh(\chi -\theta
)}Z(\theta )d\theta ,\,\,\,\,\,Z(\theta )\Omega =0\,\,  \label{Z}
\end{equation}
\begin{eqnarray}
Z(\theta _{1})Z(\theta _{2}) &=&S_{Z,Z}(\theta _{1}-\theta _{2})Z(\theta
_{2})Z(\theta _{1})  \label{com} \\
Z(\theta _{1})Z^{\ast }(\theta _{2}) &=&\delta (\theta _{1}-\theta _{2})%
\mathbf{1}+S_{Z,Z^{\ast }}(\theta _{1}-\theta _{2})Z^{\ast }(\theta
_{2})Z(\theta _{1})  \notag
\end{eqnarray}
For the moment the $S^{^{\prime }}s$ are simply Lorentz-covariant (only
rapidity differences appear) functions which for algebraic consistency
fulfil unitarity $\overline{S(\theta )}=S(-\theta ).$ We assume (for
simplicity) that the state space contains only one type of particle.

Before continuing with the special situation we introduce a useful general
definitions.

\begin{definition}
A field operator F(x) is called ``one-particle \textbf{p}olarization \textbf{%
f}ree'' if F(x)$\Omega $ and F$^{\ast }$(x)$\Omega $ have only one-particle
components (for any one of the irreducible particle spaces in the theory)
\end{definition}

For polarization free $F(x)^{\prime }s$ the vector $F^{\#}(x)\Omega $ is on
mass-shell i.e. has a Fourier transform in terms of $Z^{\ast }(\theta
)\Omega $, with $Z(\theta )\Omega =0.$ Note that the definition does not yet
require that $F(x)$ itself to be on-shell. We are however interested in $%
F(x)^{\prime }s$ which upon smearing with test functions restricted to a
subspace $\mathcal{L}$ generate algebras 
\begin{equation}
\mathcal{A}=alg\left\{ F(\hat{f})=\int F(x)\hat{f}(x)d^{d}x\mid \hat{f}\in 
\mathcal{L}\right\}
\end{equation}
which on the one hand are big enough in order to create a dense set of
states if applied to $\Omega ,$ but on the other hand allow for an equally
big commutant algebra $\mathcal{A}^{\prime },$ in short the PF's should
generate an $\mathcal{A}$ which is cyclic and separating with respect to the
vacuum. As a result of $F(\hat{f})A^{\prime }\Omega =A^{\prime }F(\hat{f}%
)\Omega $ for $A^{\prime }\in \mathcal{A}^{\prime },$ the on-shell aspect of
the vectors is transferred to the operators, i.e. formula (\ref{Z}) for $%
F(x) $ is valid. The $\mathcal{L}^{\prime }s$ we have in mind are subspaces
of localized test functions $\mathcal{L}=\left\{ \hat{f}\mid supp\hat{f}%
\subset \mathcal{O}\right\} $. But as a consequence of an old theorem by
Jost and the present author \cite{St-Wigh}, this immediately limits the
admissable localization properties. If the field is pointlike local, this
theorem forces the $F$ to be a free field, and by a slight massaging of the
proof this would continue to hold for $F^{\prime }s$ which have a compact
Minkowski space localization. Even for noncompact localizations which are
properly contained in a wedge (i.e. a Lorentz transformed of the standard
wedge $x_{1}>\left| x_{0}\right| )$ this clash with interactions continues%
\footnote{%
I owe this general model independent insight to D. Buchholz, private remark,
unpublished.} and the only consistent value of the $S$-functions in the
above Ansatz are $S=\pm 1$ i.e. free Bosons/Fermions. The smallest region
for which these arguments break down are full wedges. The following theorem
shows that indeed wedge localization in d=1+1 is consistent with nontrivial
interactions and the result emerging from the above Ansatz in formula (\ref
{com}) is quite surprising.

One finds that the coefficients are related to each other and fulfil the
complete Zamolodchikov-Faddeev algebra if and only if the $F(\hat{f}%
)^{\prime }s$ with $supp\hat{f}\in W$ generate wedge localized algebra, thus
unraveling the physical significance of this formally introduced algebraic
structure in terms of wedge localization \cite{Sch}\cite{S-W1}.

This is not the first time in physics that wedges play a prominent role. In
Unruh's Minkowski space illustration of the origin of thermal aspects of
quantum matter encapsulated behind a horizon, in the first application of
Tomita's modular theory by Bisognano and Wichmann and now in the inverse use
of the Bisognano-Wichmann theorem for the direct construction of local
algebras, in all cases one encounters the fundamental role of wedge
localization and wedge algebras. In the present case we find \cite{SchAOP} 
\cite{S-W1}\cite{SchGoe}

\begin{proposition}
The requirement of wedge localization of a PF operator $F(f)=\int F(x)\hat{f}%
(x)d^{2}x,\,suppf\in W$ with $F$ fulfilling formula \ref{com} is equivalent
to the Zamolodchikov-Faddeev structure of the Z-algebra. In particular the
thermal (Hawking-Unruh) KMS condition on their Wightman correlation
functions correspond to the crossing symmetry of the $S$-coefficient
functions. The corresponding F's cannot be localized in smaller regions i.e.
the localization of F(\^{f}) with supp\^{f}$\in \mathcal{O}\subset W$ is not
in $\mathcal{O}$ but still uses all of $W.$
\end{proposition}

The reader can find the proof which amounts to a simple computation in \cite
{S-W1}\cite{SchGoe}. Of course the F's are not ordinary (Lagrangian or
Wightman-) fields, since there localization does not follow the decreasing
support properties of f's inside the wedge and therefore F(f) is the better
notation than F(x). Since \textbf{p}olarization \textbf{f}ree \textbf{g}%
enerators $F$ will only play a role as wedge generators, we will simply use
the abbreviation \textbf{PFG} standing for
``polarization-free-wedge-generators''.

A moments thinking about the special situation reveals that the modular
structure, i.e. the existence of the antilinear unbounded Tomita involution $%
S_{T}$ (the subscript serves to distinguish this time-honored modular
notation from the equally time-honored notation for the scattering operator)
is the general cause underlying the above observation. In fact the modular
``basic law'' for the physical wedge algebra is: 
\begin{equation}
S_{T}A\Omega =A^{\ast }\Omega ,\,\,\,A\in \mathcal{A}(W)
\end{equation}
which defines the antilinear, unbounded, closable, involutive (on its
domain) Tomita operator $S_{T}.$ Its polar decomposition 
\begin{equation}
S_{T}=J\Delta ^{\frac{1}{2}}
\end{equation}
defines a positive unbounded $\Delta ^{\frac{1}{2}}$ and an antiunitary
involutive $J$ and the nontrivial part of Tomita's theorem (with
improvements by Takesaki) is that the unitaty $\Delta ^{it}$ implements an
automorphism of the algebra i.e. $\sigma _{t}(\mathcal{A})\equiv \Delta ^{it}%
\mathcal{A}\Delta ^{-it}=\mathcal{A}$ and the $J$ maps into antiunitarily
into its commutant $j(\mathcal{A})\equiv J\mathcal{A}J=\mathcal{A}^{\prime
}. $ For the case at hand (the Bisognano-Wichmann situation) these operators
have their following physical aliases: 
\begin{eqnarray}
\Delta ^{it} &=&U(\Lambda (-2\pi t))=U_{in}(\Lambda (-2\pi t)) \\
J &=&SJ_{in}  \notag
\end{eqnarray}
where $U_{in}(\cdot )$ is the unitary representation of the
Poincar\'{e}-group in the incoming Fock space and $J(J_{in})$ is the TCP
operator (its free field incoming version). The last relation shows clearly
that the S-matrix is a relative modular invariant of the wedge algebra.

The wedge situation is a special illustration for the Tomita theory covered
by the Bisognano-Wichmann theorem \cite{Ha}. In that case both operators
have well known physical aliases; the modular group is the one-parametric
wedge affiliated Lorentz boost group $\Delta ^{it}=U(\Lambda (-2\pi t),$ and
the $J$ in $d=1+1$ LQP's is the fundamental TCP-operator as derived from
first principles by R. Jost \cite{Ha}; in higher dimensions it is only
different from TCP by a $\pi $-rotation around the spatial wedge axis. The
formula for the modular operator in terms of \ the scattering matrix (which
contains the information about the interaction) is not part of that theorem
and as such is new. However it turns out to be just a modular adapted
transcription of the TCP transformation law of the textbooks \cite{St-Wigh}.
The prerequisite for the general Tomita situation is that the vector in the
pair \{algebra, reference vector\} is cyclic and separating i.e. there is no
annihilation operators in the von Neumann algebra or equivalently: its
commutant is cyclic relative to the reference vector. In LQP these
properties are guarantied for localization regions $\mathcal{O}$ with
nontrivial causal complement $\mathcal{O}^{\prime }$ thanks to the
Reeh-Schlieder theorem. In terms of the correlation functions of the
generators, the wedge localization affiliation of the generated algebra is
nothing but the KMS condition (which is checked in the above mentioned proof 
\cite{S-W}).

The construction of the local QFT behind the S-matrix of the above model is
of course not finished with that of its wedge algebra. The essential next
step is the construction of its double cone algebras\footnote{%
The sharpening of localization via algebraic intersections is the essential
difference to usual QFT even including Wightman's approach. I was quite
surprized when in this way I obtained the same recursion formula \cite{Sch}
which in Smirnov's ``axiomatic'' approach \cite{Smi} (not the usual QFT
axioms but rather some calculational recipes designed for factorizing
models).} via the demonstration of the nontriviality ($\neq C\cdot \mathbf{1}
$) of the intersection of the right wedge algebra with its translated
opposite left wedge. It is precisely here where the idea of holography
enters the game. It is much easier to show the nontriviality of the
holographic image of this situation.

The crucial idea is to look at the relative commutant for light-like
translations for say $a_{+}=(1,1)$ 
\begin{equation}
\mathcal{A}(W_{+})^{\prime }\cap \mathcal{A}(W)
\end{equation}
where $\mathcal{A}(W_{+})$ is the $a_{+}$-shifted wedge algebra. $\mathcal{A}%
(W_{+})\subset $ $\mathcal{A}(W)$ is almost a modular inclusion, i.e. the
modular group of $\mathcal{A}(W)$ i.e. the Lorentz-boost in one direction
acts on $\mathcal{A}(W_{+})$ as a compression into itself. The only missing
property is the standardness of the relative commutant in $H$ with respect
to $\Omega .$ But this is easily achieved by projection onto the cyclicity
space $\mathcal{M}_{+}\Omega $ 
\begin{equation}
H_{+}=P_{+}H\subset H=\overline{\mathcal{A}(W)\Omega }
\end{equation}
Using a theorem of Takesaki, the reduced inclusion defines again a modular
inclusion in its own right from which one may reconstruct a positive energy
translation $\hat{U}(a)$ which then can be used to define a reduced net
indexed by intervals 
\begin{eqnarray}
\mathcal{A}(I_{a,e^{2\pi t}+a}) &=&\hat{U}(a)\Delta ^{-it}E_{+}\left( 
\mathcal{A}(W_{a_{+}})^{\prime }\cap \mathcal{A}(W)\right) \Delta ^{it}\hat{U%
}^{-1}(a,a) \\
\mathcal{M}_{+} &\equiv &\cup _{t}A(I_{0,e^{2\pi t}}),\,\,E_{+}(\mathcal{A}%
(W))=\mathcal{M}_{+}=P_{+}\mathcal{A}(W)P_{+}  \notag
\end{eqnarray}
The reduced net can be shown to be ``standard'' and the set of standard
modular inclusion is known to be isomorphic to the set of all chiral
conformal field theories (S-W). Therefore each d=1+1 net comes associated
with a ``satellite'' chiral conformal net. This is the rigorous modular
version of \ holography and again, as in the AdS case treated by Rehren,
this association is outside the range of Lagrangian quantization since there
is no Lagrangian or euclidean functional integral process which can describe
properly this transmutation of degrees of freedom. At the time of writing of
this essay, the computations for the existence proof of the factorizing
models ($\sim $double cone algebra nontriviality) have not been finished.

The use of the holography idea for higher dimensional QFT's is more
involved. If one carries out the previous modular inclusion construction,
one realizes that, because of the transversal indeterminacy of the chiral
conformal theory attached by modular inclusion (which is localization-wise
really attached to a whole light front rather than a light ray), the chiral
theory is, contrary to Rehren's AdS treatment, not yet sufficient for a
reconstruction of the original theory from the holographic image. It turns
out that by tilting the Lorentz-boost of the original wedge around one of
its light rays one generates a ''stalk'' of conformal theories which, more
analogous to a scanning process than holography, allows the reconstruction 
\cite{S-W2} (called ``blow-up'' in the paper) of the full net theory. It
uses an enrichment of modular inclusion, called modular intersections, which
in its geometrical interpretation corresponds to the interaction of two
different wedges which have one light ray in common. Modular inclusions and
modular intersections constitute presently the most powerful
conceptual/mathematical instruments which LQP has to offer.

It should be clear to those who know a bit about the two-dimensional
bootstrap-formfactor constructive program, that the two-dimensional modular
method for factorizing models (i.e. those which are defined by a
matrix-generalization of the Ansatz at the beginning of this section) is
what lies behind the formfactor program initiated by Karowski-Weisz and
Smirnov \cite{K-W}\cite{Smi}. It does however more than only justify those
very successful collection of non-Lagrangian cooking recipes, in that it
promises to solve the difficult and not fully understood problem of the
correlation functions of pointlike quantum fields (using again standard QFT
concepts) in those models. In fact this difficulty, known to every expert of
the bootstrap-formfactor program in the conventional setting results from
the fact that there is no natural basis in interacting field space as the
Wick composites for free fields. Therefore it is better to avoid fields
altogether and characterize the physical content of a theory in terms of its
basis independent double cone algebras. In their nontriviality demonstration
as well as in their actual construction, the holography, as we have argued,
plays an essential role.

Since Chew's S-matrix bootstrap program (i.e. the formulation of the
nonlinear S-matrix axioms as well as the actual construction of interesting
examples) only works for the d=1+1 factorizing models, there is no hope to
do higher dimensional QFT with modular methods in a two-step process of
calculating first $S$ and then the associated wedge algebra. Rather one has
to understand the structure of correlation functions of the wedge generators 
$F(f)$ (which turn out to be uniquely fixed in terms of $S)$ and of $S$
itself simultaneous. This is presently only imaginable in a perturbative
spirit. But note that this would be a perturbative approach for wedge
algebras and not for individual fields, i.e. technically speaking for the
whole space of formfactors generated from the PFG's sandwiched between
incoming particle ket vectors and outgoing bra vectors without any natural
way of distinguishing individual elements \cite{SchGoe}. Refinements and
distinctions have to go via improvements of localizations, which in modular
theory can only happen through algebraic intersections.

In order to return at the end of my essay to the Wichmann's S-matrix research%
\footnote{%
I do not know whether Wichmann, while working on wedge localization and
modular theory, was aware about these strong connections with his previous
S-matrix research. It may have been another instance of the role the
subconcious on scientific research.} carried out at the beginning of his
research carrier at Berkeley, I would like to use some recent personal
experience of my own as a vehicle to recapture some flavor of those times.

Shortly after string theorist picked the big Latin letter ``M'' for one of
their recent inventions, but before the much clearer AdS proposal (note the
small d there!) attracted the attention, I was struck by the wealth of
coincidences of some of the string theoretic statements, especially in
connection with transmutations and counting of degrees of freedom, light
cone physics (in particular their Galilean group affiliated with the light
cone) and ideas on holography, with recent results about consequences of
modular theory. Although I admittedly do not understand string theory from a
physical point of view, I do think (most of my colleagues from algebraic QFT
do not share such optimistic ideas) that the used differential-geometric
quantization formalism, together with the relics of physical locality and
spectral properties which such a generalized standard formalism inexorably
contains and which are (even in in arguing along differential geometric
instead of local quantum physical lines) hard to loose, constitutes a
powerful mathematical machinery for the discovery of new structures; even
though physical interpretations which could reconcile string theory with the
physically (but not mathematically!) more conservative local quantum physics
are hard to see, and even when faintly visible, probably not always correct.
Apparently the string formalism uses (to me hidden) ideas which carries
string physicists beyond the confines of possibilities allowed by Lagrangian
quantization, and in this way achieves similar ``miracles'' as modular
theory (which originated by faithfulness to all the principles underlying
standard QFT, but not its formalism).

But my proposal to include ``modular'' (in the sense of d=1+3 LQP without
``curling up'' unwanted dimensions by semiclassical Klein-Kaluza ideas) also
in the list of possible interpretations of the letter ``M'' did not find the
approval of the referee who claimed that all this is accidental and spurious
(whereupon I withdrew my admittedly rather speculative notes from the hep-th
server).

Fortunately the arrival of the clearer AdS structures has made it possible
to have at least some realistic comparisons\cite{BFS}\cite{Re} if not
directly of string theory, then at least of some of what are believed to be
consequences of its underlying philosophy.

My attitude towards the issue string/modular theory\footnote{%
I still believe that both string theory and LQP leave the rather narrow
confines of standard Lagrangian QFT, but for different reasons and with
different aims. Whereas for the followers of string theory, QFT was
identified with the standard text book Lagrangian or functional
quantization, and therefore the (revolutionary) departure happened on the
physical side by keeping as much as possible of the standard formalism, LQP
is totally conservative with respect to the underlying physical principles,
but revolutionary on their mathematical and conceptual implementations.
String theory, after its second revolution, wanted to be (or at least to
incorporate) quantum gravity, whereas LQP definitely wants to stay with
laboratory physics. A very instructive illustration of this difference is
supplied by comparing Witten's approach \cite{Wi} to AdS versus that in \cite
{Re} \cite{BFS}.} did not change, in part due to the fact that I have much
deeper, almost archaic reasons, which probably also the Jubilar from his
early Berkeley S-matrix days can share with me. I am referring to the
mysterious role of crossing symmetry which at the same time was (apart from
unitarity) by far the physically most important input into the Berkeley
(primarily Chew's) S-matrix bootstrap program; the analyticity, to the
extend that it was not needed in the formulation of crossing, was more of a
technical nature. Now, with all the hindsight of the distance in time and
significant advances about the consequences of the Jubilar's modular
contributions, it is becoming gradually clear that those two topics belong
together, and that there could not have been conceptual progress on crossing
symmetry without a comprehensive modular understanding of the wedge
situation.

Let us follow the flow of history on crossing symmetry a bit more.

In order to lift some of its mystery, as everybody of that generation
remembers, Veneziano invented the dual model and Virasoro observed
subsequently that the (onshell) S-matrix (still without its unitarity
corrections) permitted the mathematical trick of a representation in terms
of a lower dimensional (offshell) QFT. This was the birth of string theory,
never mind its several revolutions and semantic changes which happened in
the course of its conversion of an original nonperturbative proposal for a
strong interacting S-matrix\footnote{%
An S-matrix with an infinite tower of particles in a finite range of mass is
of course not compatible with a reasonable phase space behavior of quantum
physics (Hagedorn temperature and worse), but as in Feynman's perturbation
theory one would expect that from genus g=2 on, the tower (except a finite
number of particles) would transmute into second Riemann sheet resonances.
According to the best of my knowledge there is no known property of QFT
which prohibits this. But in this case, what means ``stringyness'' versus
QFT behavior?} into a TOE including quantum gravity.

Unfortunately the full understanding of the notoriously difficult crossing
symmetry (which most people thought of as an onshell imprint of the offshell
Einstein causality), and whose unraveling was worth any effort, was not
obtained. Regrettably the birth of string theory was for many especially
young theoreticians, also the point of departure into the physical blue
yonder with little chance to return.

Now with the patient and precise early work of the Jubilar on modular theory
bearing many fruits, and with the importance of crossing symmetry at the
cradle of string theory on ones mind, the proposal to occupy part of the
physically underpopulated M-universe with m as in modular theory, may after
all not turn out to be as outrageous as it appears on first sight.

\begin{acknowledgement}
I am indebted to Detlev Buchholz for giving some advice in the writing of
this manuscript and to my FU-Berlin Experimentalphysik Kollege William
Brewer, who recreated to me (over some drinks at a bar in Copacabana) a bit
of the Zeitgeist of the 60$^{ies}$ at Berkeley, when he took theory courses
from Prof. Eyvind Wichmann. Finally I thank Marinus Winnink for having
confirmed the historical correctness of some of my remarks concerning the
dawn of modular theory.
\end{acknowledgement}

\end{document}